\def\spose#1{\hbox to 0pt{#1\hss}}

\def\multleft#1{\hbox to size{\vbox {\halign {\lft{##}\cr #1}}\hfill}\par}
\def\multright#1{\hbox to size{\vbox {\halign {\rt{##}\cr #1}}\hfill}\par}

\def\today{\ifcase\month\or January\or February\or March\or April\or May\or
      June\or July\or August\or September\or October\or November\or December\fi
      \space\number\day, \number\year}
\def\s{\hbox{\phantom{5}}}	

\def\cm{{\rm\thinspace cm}}

\def\km{{\rm\thinspace km}}
\def\kpc{{\rm\thinspace kpc}}

\def\Mpc{{\rm\thinspace Mpc}}
\def\Msun{\hbox{$\rm\thinspace M_{\odot}$}}

\def\s{{\rm\thinspace s}}

\def\kmps{\hbox{$\km\s^{-1}\,$}}

\def\psqcm{\hbox{$\cm^{-2}\,$}}

\def\kmpspMpc{\hbox{$\kmps\Mpc^{-1}$}}

\def\H2{\hbox{H$_{2}$}}
\def\Lya{Ly$\alpha$}

\documentclass[usegraphicx]{mn2e}

\usepackage{graphicx}
\usepackage{times}
\usepackage{amssymb}
\include{defn}

\begin{document}
\hsize=6truein

\title[The association between gas and galaxies II: The 2-point correlation function]
{The association between gas and galaxies II: The 2-point correlation function}
\author[R.J.~Wilman et al.]
{\parbox[]{6.in} {R.J.~Wilman$^{1}$, S.L.~Morris$^{1}$, B.T.~Jannuzi$^{2}$, R.~Dav\'{e}$^{3}$, A.M.~Shone$^{1}$ \\ \\
\footnotesize
1. Department of Physics, University of Durham, South Road, Durham, DH1 3LE. \\
2. National Optical Astronomy Observatory, PO Box 26732, Tucson, AZ 85726-6732, USA. \\ 
3. University of Astronomy, Tucson, AZ 85721, USA.}}
\maketitle

\begin{abstract}
We measure the 2-point correlation function, $\xi_{\rm{AG}}$, between galaxies and quasar absorption line systems at $z<1$, 
using the dataset of Morris \& Jannuzi (2006) on 16 lines-of-sight (LOS) with UV spectroscopy and galaxy multi-object 
spectroscopy (paper I). The measurements are made in 2D redshift space out to $\pi = 20 h^{-1}$\Mpc~(comoving) along the LOS and out to 
$\sigma = 2 h^{-1}$\Mpc~projected; as a function of HI column density in the range 
$N_{\rm{HI}}=10^{13}-10^{19}$\psqcm, also for CIV absorption systems, and as a function of galaxy spectral type. This extends the 
absorber-galaxy pair analysis of paper I. We find that the amplitude of the peak in $\xi_{\rm{AG}}$ at the smallest separations increases slowly as the lower limit on $N_{\rm{HI}}$ is increased from $10^{13}$ to $10^{16}$\psqcm, and then jumps sharply (albeit 
with substantial uncertainties) for $N_{\rm{HI}} > 10^{17}$\psqcm. For CIV absorbers, the peak strength of $\xi_{\rm{AG}}$ is roughly comparable to that of HI absorbers with $N_{\rm{HI}} > 10^{16.5}$\psqcm, consistent with the finding that the CIV absorbers are associated with strong HI absorbers.

We do not reproduce the differences reported by Chen et al. between 1D
$\xi_{\rm{AG}}$ measurements using galaxy sub-samples of different
spectral types. However, the full impact on the measurements of systematic differences in our
samples is hard to quantify. We compare the
observations with smoothed particle hydrodynamical (SPH) simulations
and discover that in the observations
$\xi_{\rm{AG}}$ is more concentrated to the smallest separations than
in the simulations. The latter also display a `finger of god'
elongation of $\xi_{\rm{AG}}$ along the line of sight in redshift
space, which is absent from our data, but similar to that found by Ryan-Weber for the cross-correlation of quasar absorbers and HI-emission-selected galaxies. The physical origin of these `fingers of god' is unclear and we thus highlight several possible areas for further investigation. 


\end{abstract}


\section{INTRODUCTION}
We are investigating the nature and extent of the relationship between gas, as probed by neutral hydrogen quasar absorption line systems, 
and the luminous baryonic matter traced by galaxies at cosmic epochs $z<1$. In this paper we extend the work presented by Morris \& Jannuzi~(2006) (paper I), where we set out the motivation for the project, defined the 16 lines-of-sight (LOS) observational dataset, and established the existence of an absorber--galaxy correlation using pair count analysis. We extend the analysis of this dataset 
to compute the absorber-galaxy 2-point correlation function, $\xi_{\rm{AG}}$, as a function of HI column density and compare the results with other measurements and simulations.

As discussed and referenced extensively in paper I, the history of work in this area has been characterised by a debate between two
 extreme positions: (i) that low-redshift Ly$\alpha$ absorbers arise in the haloes of luminous galaxies; or (ii) that they are 
part of a filamentary network pervading the intergalactic medium (IGM), and are related to galaxies only to the extent that they both trace 
the underlying dark matter distribution. This polarised picture is undoubtedly a simplification 
of a reality which now seems likely to require a mixture of the two components in proportions which vary as a function of galaxy type, 
absorber column density and cosmic epoch. A further complication arises from the expectation that at high 
redshift galaxies are expected to blow most of their star-forming material out into the IGM as ``superwinds''. Such outflows are 
hypothesised to pollute the IGM with metals (Aguirre et al.~2001), to prevent the formation of over-luminous galaxies at $z=0$ (Benson et al.~2003) and to account for the low-fraction of $z=0$ baryons locked in stars. The measurement of $\xi_{\rm{AG}}$ provides a method to 
characterise these feedback processes and to facilitate a direct comparison with simulations. At $z \sim 2-3$, attempts have been made to measure the impact of superwinds on the IGM by cross-correlating HI and CIV absorption systems with UV selected galaxies, but the early results were 
ambiguous (Adelberger et al.~2003). Subsequent measurements have clarified the situation (Adelberger et al.~2005), and reveal that strong 
Ly$\alpha$ absorption is produced in intergalactic gas within $1 h^{-1}$ comoving \Mpc~of $\sim 2/3$ of such galaxies, and that strong CIV absorption ($N_{\rm{CIV}} \gg 10^{14}$\psqcm) extends out to impact parameters of 40\kpc~(proper) with a velocity range $\Delta v > 260$\kmps. Whether such superwind events have imprinted a relic signature on the galaxy-IGM relationship which persists 
to $z<1$ is not known, but the measurements presented in the present paper may provide some contraints on this issue (see also Rauch et al~2005).

In paper I we showed that absorbers and galaxies are correlated out to impact parameters of at least 1.5\Mpc~(physical) but that this 
clustering is weaker than the galaxy-galaxy clustering in the sample. The velocity differences between absorber--galaxy pairs 
with detected CIV absorption are typically smaller than those with HI absorption only, but variations with HI column density were not 
assessed. Several other works have cross-correlated absorption line catalogues with various low-z galaxy samples and reached broadly similar,
but not identical, conclusions. Ryan-Weber~(2006) cross-correlated galaxies selected in HI 21\cm~emission with low column density ($N_{\rm{HI}} 
< 10^{15}$\psqcm) Ly$\alpha$ absorbers and found the clustering to be similar in strength to the clustering of the galaxies in the sample in redshift 
space, but somewhat stronger in real space on scales of 1--10$h^{-1}$\Mpc. At somewhat higher column density, Bouch\'{e} et al.~(2004) 
cross-correlated MgII absorbers with luminous red galaxies (LRGs) at $0.4 \leq z \leq 0.8$, and over comoving scales of 
0.05--13$h^{-1}$\Mpc~found that the MgII-LRG cross-correlation amplitude is $\simeq 30$ per cent weaker than the LRG-LRG clustering. In a 
series of papers Prochaska and collaborators have investigated the clustering of $z<0.5$ 
galaxies around absorbers along the sightline to the quasar PKS 0405-123 (Prochaska et al.~2004; Chen et al.~2005; Prochaska et al.~2006). 
To the extent that it is possible using a single sightline, they showed that emission-line galaxies and absorbers with $N_{\rm{HI}} 
> 10^{14}$\psqcm~cluster with a strength comparable to the galaxy-galaxy clustering, whilst the absorption-line galaxies in their sample 
are not correlated with absorbers \footnote{For conformity with earlier works, we retain the nomenclature of `emission-line' and `absorption-line' galaxies to refer to galaxies whose spectra are dominated by emission line and absorption line features, respectively. The latter should not be confused with quasar absorption lines against which the galaxy populations are cross-correlated.}. They also showed that absorbers with $N_{\rm{HI}} < 10^{13.6}$\psqcm~are distributed more randomly with
respect to galaxies, and that the strength of $\xi_{\rm{AG}}$ is insensitive to $N_{\rm{HI}}$ for $10^{13.6}  < N_{\rm{HI}} < 10^{16.5}$\psqcm. Also for this sightline, Williger et al.~(2006) found that the absorber-galaxy correlation function is significant out to $\Delta v < 250$\kmps~and grows with minimum absorber column density. Some of these conclusions can be tested with our multiple LOS dataset and that is our aim here.

For completeness, we note that Chen et al. (2001) presented a sample of quasar lines of sight with HST FOS observations together with galaxy
redshifts. As mentioned in Morris \& Jannuzi~(2006), we have chosen not to fold their data into our current analysis due to significant
differences in survey strategy.

Throughout the paper we adopt the cosmological parameters $\Omega_{\rm{M}}=0.3$, $\Omega_{\rm{\Lambda}}=0.7$, and $H_{\rm{0}}=100h$\kmpspMpc.

\section{THE ABSORBER-GALAXY 2-POINT CORRELATION FUNCTION}

The absorber galaxy cross correlation function, $\xi_{\rm{AG}}$, is defined through the expression for the conditional probability $dP$ of finding a galaxy in a volume $dV$ at position $\bf{r_{2}}$, given that there is an absorber at position $\bf{r_{1}}$:

\begin{equation}
dP= n dV [1 + \xi_{\rm{AG}} (\bf{r_{2} - r_{1} })],
\end{equation}
where $n$ is the unconditional galaxy density.

The samples of absorption lines and galaxies which we will use to compute $\xi_{\rm{AG}}$ are identical
to those defined in paper I, where full details can be found. In brief, the dataset comprises 16 quasar fields (from the {\em Hubble Space Telescope} Quasar Absorption Line Key Project dataset; Jannuzi et al.~1998), whose LOS contain a total of 381 \Lya~absorption lines and 30 CIV systems (25 at $z <1$), and a total of 685 galaxies with redshifts $z<1$. As discussed in paper I, 49 of these galaxies have recession velocities $<500$\kmps~and are most likely stars, but their exact identification is irrelevant for this study because 98 per cent of the HI absorbers are in any case at $z>0.3$. The number of galaxies in the redshift range $0.3 \leq z \leq 1$ which effectively overlap with the absorption line sample is 379.

The following estimator, introduced by 
Davis \& Peebles~(1983), is used to compute $\xi_{\rm{AG}}$ as a function of projected separation, $\sigma$, and line-of-sight separation, $\pi$:

\begin{equation}
 \xi_{\rm{AG}}(\sigma,\pi) = AG/AR - 1,
\end{equation}

where AG is the number of real absorber-galaxy pairs in the region $(\sigma -\delta\sigma/2:\sigma+\delta\sigma/2,\pi -\delta\pi/2:\pi + \delta\pi/2)$ and AR is the suitably normalized number of absorber-random galaxy pairs at the same location. Following Bouch\'{e} et al.~(2004), the numbers of pairs in each of the quasar fields are normalized separately such that $AR = \Sigma AR^{i} N_{\rm{g}}^{\rm{i}}/N_{\rm{r}}^{\rm{i}}$, where the summation is over the 16 fields and $N_{\rm{g}}^{\rm{i}}$ and $N_{\rm{r}}^{\rm{i}}$ are the numbers of real and random galaxies in field $i$, respectively. All absorber-galaxy separations are computed in comoving redshift space and the absorbers are assumed to be located at the sky position of the quasar LOS. For the assumed flat cosmology, the comoving distance to an object at redshift $z$ is given by:

\begin{equation}
X_{\rm{CM}}(z) =  \frac{c}{H_{\rm{0}}} \int_0^{z} \frac{dz}{\sqrt{\Omega_{\rm{\Lambda}}  + \Omega_{\rm{M}}(1 + z)^{3}}}
\end{equation}
and hence for a close absorber-galaxy pair separated by $\Delta z$ in redshift 
and by an angle $\Delta \theta$ on the sky, $\sigma$ and $\pi$ are given by:

\begin{equation}
\sigma = X_{\rm{CM}} \Delta \theta
\end{equation}
and
\begin{equation}
\pi = \frac{dX_{\rm{CM}}}{dz} \Delta z.
\end{equation}

Several crucial details of the computational method, namely the generation of the random galaxy catalogue, the binning and smoothing of $\xi_{\rm{AG}}$, and the calculation of uncertainties, are dealt with in the next subsection.

\subsection{Computational method}

\subsubsection*{Random galaxy catalogues}
Our method requires a separate random galaxy catalogue for each quasar field in order to reflect any field-to-field differences in the galaxy 
selection function. The latter mainly arise due to the placement of slits and not from differences in spectroscopic depth. Following 
Adelberger et al.~(2003), we form such catalogues by replacing each real galaxy with 100 random galaxies, whose redshifts are drawn at random from the galaxy redshift distribution of the other fields, i.e. the real and random galaxies have the same positions on the sky, but different redshifts. This reflects the fact that galaxies in the sample must be located at the position of the slits.

\subsubsection*{Spatial binning and smoothing of $\xi_{\rm{AG}}$}
To reduce the effects of shot noise due to the small numbers of absorber-galaxy pairs on the measurement of $\xi_{\rm{AG}}$, some consideration must be given to the form of the numerical coordinate grid, $(\sigma, \pi)$, on which it is evaluated.  There are two approaches. The
first approach adopts a grid whose cell-size is large enough to include a sufficient number of absorber-galaxy pairs. An adaptively-sized mesh would be
the best way to handle strong variations in the density of absorber-galaxy pairs, but this would have the tendency to dilute the signal in 
regions where the paucity of pairs arises from a genuinely strong anticorrelation signal with $\xi_{\rm{AG}} \simeq -1$. 

An alternative approach is to bin the measurements much more finely (if at all) and then to smooth the resulting field to remove the noise. 
Again, an adaptively-sized smoothing kernel could be the best option. The two underlying density fields, $AG(\sigma,\pi)$ and 
$AR(\sigma,\pi)$, must be smoothed first and then ratioed to obtain the smoothed $\xi_{\rm{AG}}$, as in eqn.~(2). 

In this work we present results obtained using both the above approaches: (i) on a uniform grid sampled as $\sigma=0,0.4,0.8...2 h^{-1}$\Mpc, $\pi = 2,4,6,...20 h^{-1}$\Mpc-- referred to as the `binned' results (the 2$h^{-1}$\Mpc~bin size in the $\pi$ direction reflects the $\simeq 200$\kmps~spectral resolution; in the $\sigma$ direction, the field sizes dictate that galaxies are present out to maximum comoving impact parameters of $\simeq 2h^{-1}$\Mpc, which we have arbitrarily split into 5 equally spaced bins); (ii) on a finely-space grid with the underlying density fields smoothed by a gaussian of $FWHM = 0.8 \times 4 h^{-1}$\Mpc, referred to as the `smoothed' results. We did not experiment with any adaptively-sized grids or smoothing kernels.

\subsubsection*{Uncertainties on $\xi_{\rm{AG}}$} 
Using the same method as Ryan-Weber~(2006), we attempted to measure the uncertainties on $\xi_{\rm{AG}}$ using jackknife resampling in which the correlation function is computed $N=16$ times with one sightline removed each time. 

\begin{equation} 
\sigma_{\rm{\xi}}^{2} = \frac{N-1}{N} \sum_{\rm{i}} (\bar{\xi} - \xi_{\rm{i}})^{2}
\end{equation}

A disadvantage of this method is that at locations where there are no absorber-galaxy pairs in the full sample (i.e. $\xi_{\rm{AG}} = -1$), all the individual $\xi_{\rm{i}}=-1$ and hence $\sigma_{\rm{\xi}}=0$. This highlights another problem of assessing the significance of the measured  $\xi_{\rm{AG}}$ values, namely that there is no obvious figure of merit analogous to a signal-to-noise ratio, i.e. $\xi_{\rm{AG}}/\sigma_{\rm{\xi}}$ is not a straightforward indication of a correlation's significance. An simpler alternative to characterise the significance of the observed correlation is to use Possion statistics, i.e. to calculate the probability of observing the real number of absorber-galaxy pairs (AG) in a given cell, given the expected number (AR). We will utilise both these approaches.

\section{RESULTS AND COMPARISON WITH OTHER WORKS}

In so far as the sizes of our absorber and galaxy samples permit, our goal is to measure $\xi_{\rm{AG}}$ as a function of both HI column density and 
galaxy spectral type, and to compare with other results in the literature.  

\subsection{Dependence of absorber-galaxy cross-correlation function ($\xi_{\rm{AG}}$) on absorber HI column density}

The absorption line sample spans a range of HI column density from $N_{\rm{HI}} = 10^{13}-10^{19}$\psqcm, with the bulk of the lines being at the lower end of this range, as shown by the distribution in Table~1. As discussed in paper I, the modest resolution of the {\em HST} Faint Object Spectrograph (FOS) (R=1300; 230\kmps) means that the \Lya~absorption features were typically unresolved, so column densities were inferred from the equivalent width assuming a Doppler $b$-parameter of 30\kmps. As shown in Fig.~1 of Weymann et al.~(1998), the $4.5\sigma$ lower detection limit on the rest equivalent width (REW) of absorption features in these quasars decreases from 0.2\AA~at $z=0.3$ to 0.1\AA~at $z=1$. Considering that the overlap between the galaxy and absorber sample is skewed toward the lower end of this redshift range, the lower detection limit for the absorber sample is effectively $REW \simeq 0.2$\AA, equating to $N_{\rm{HI}} \simeq 10^{13.6}$\psqcm. We note, however, that the Key Project absorption line sample is not strictly complete for any REW value, but has a detection limit which varies with both LOS and redshift (as discussed at length in paper I). 

\begin{figure*}
\includegraphics[width=0.84\textwidth,angle=0]{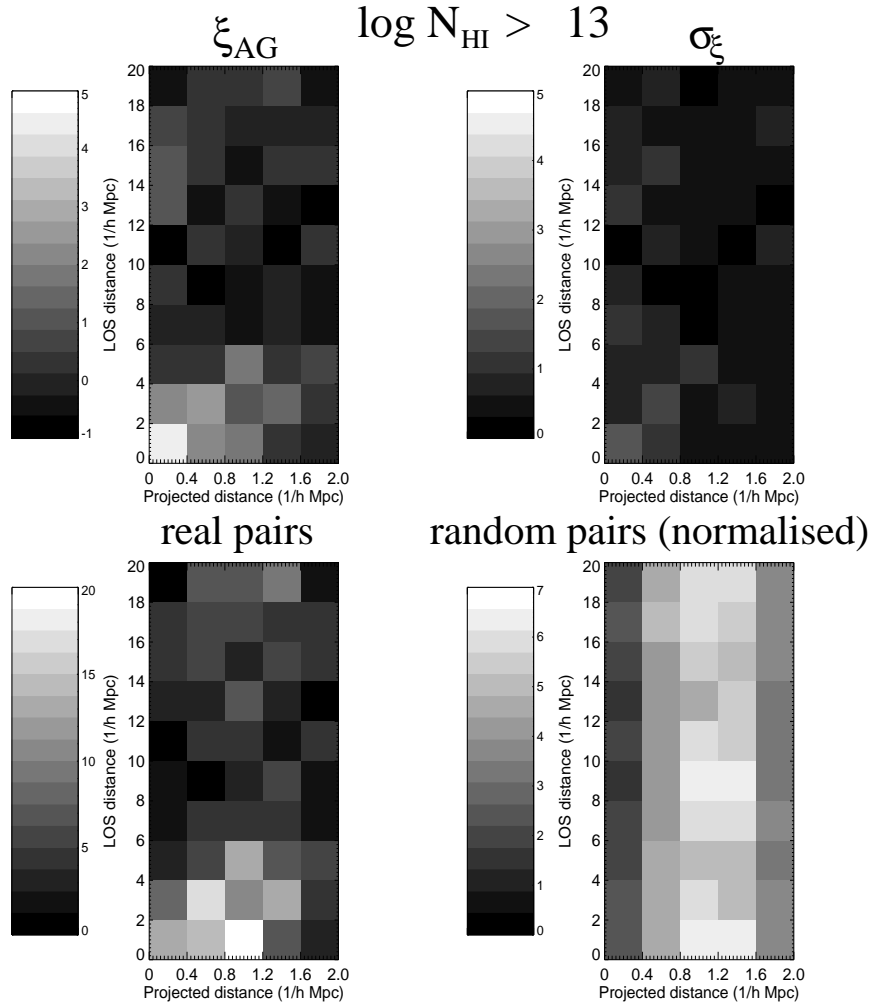}
\caption{\normalsize For the complete sample of absorbers with $N_{\rm{HI}} > 10^{13}$\psqcm, the figure shows the absorber-galaxy 2-point correlation function (top left), $\xi_{\rm{AG}}$, determined using the real and random absorber-galaxy pairs counts (AG and AR, respectively, in eqn~(2)) (lower left and right, respectively). Also shown is the uncertainty on $\xi_{\rm{AG}}$ derived using jackknife resampling, eqn.~(6) (top right).}
\label{fig:xi13bin}
\end{figure*}

In the first calculation of $\xi_{\rm{AG}}$ we use the entire absorption line sample, i.e. $N_{\rm{HI}} > 10^{13}$\psqcm. The results are shown in Fig.~\ref{fig:xi13bin} where we also display the jackknife errors $\sigma_{\rm{\xi}}$ and the distributions of absorber--galaxy real (AG) and random (AR) pairs. The vertical banding in the distribution of random pairs results from an increased volume per cell at larger projected distances and a fall-off in the galaxy density at the largest radii in our sample. Observe that whilst the correlation function itself clearly peaks in the cell at the smallest separations, the peak in the absolute number of absorber-galaxy 
pairs is spatially offset, due to volume effects. This highlights the
value of basing inferences about statistical associations on
$\xi_{\rm{AG}}$ rather than on the raw pair counts. The peak
`signal-to-noise' ratio, $\xi_{\rm{AG}}/\sigma_{\rm{\xi}}$, is
approximately 3. 

\begin{figure*}
\includegraphics[width=0.84\textwidth,angle=0]{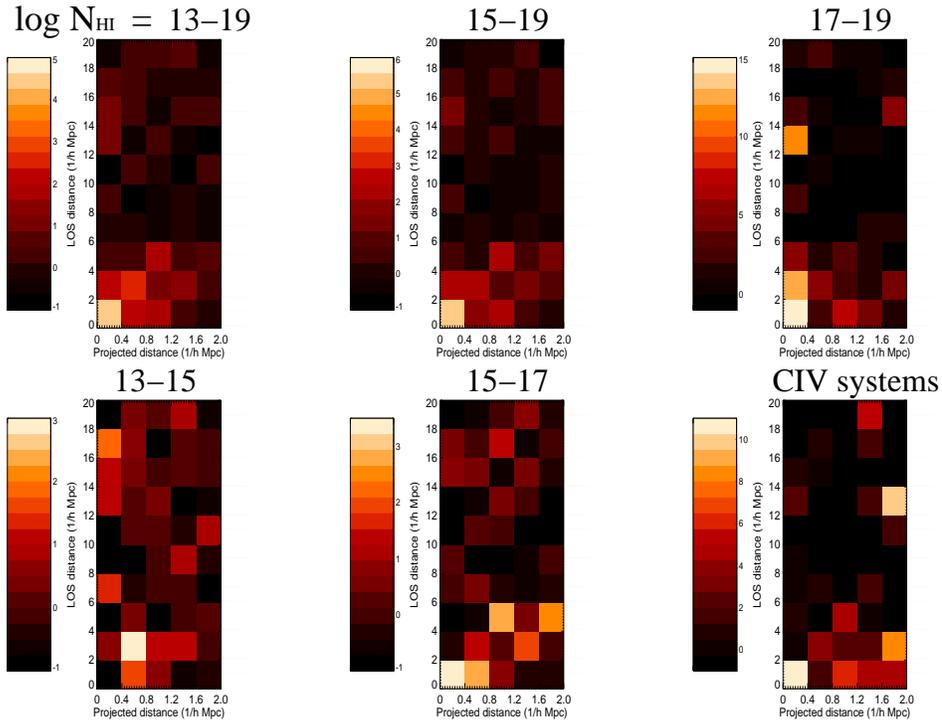}
\caption{\normalsize $\xi_{\rm{AG}}$ evaluated in 2-D redshift space
for various ranges in HI column density (as indicated) and for CIV
systems. Note the variations in the colour bar scales, and 
refer to the estimated uncertainties in Table 2 when comparing the peak
$\xi_{\rm{AG}}$ values of the subpanels.}
\label{fig:ALLbin}
\end{figure*}

\begin{figure*}
\includegraphics[width=0.84\textwidth,angle=0]{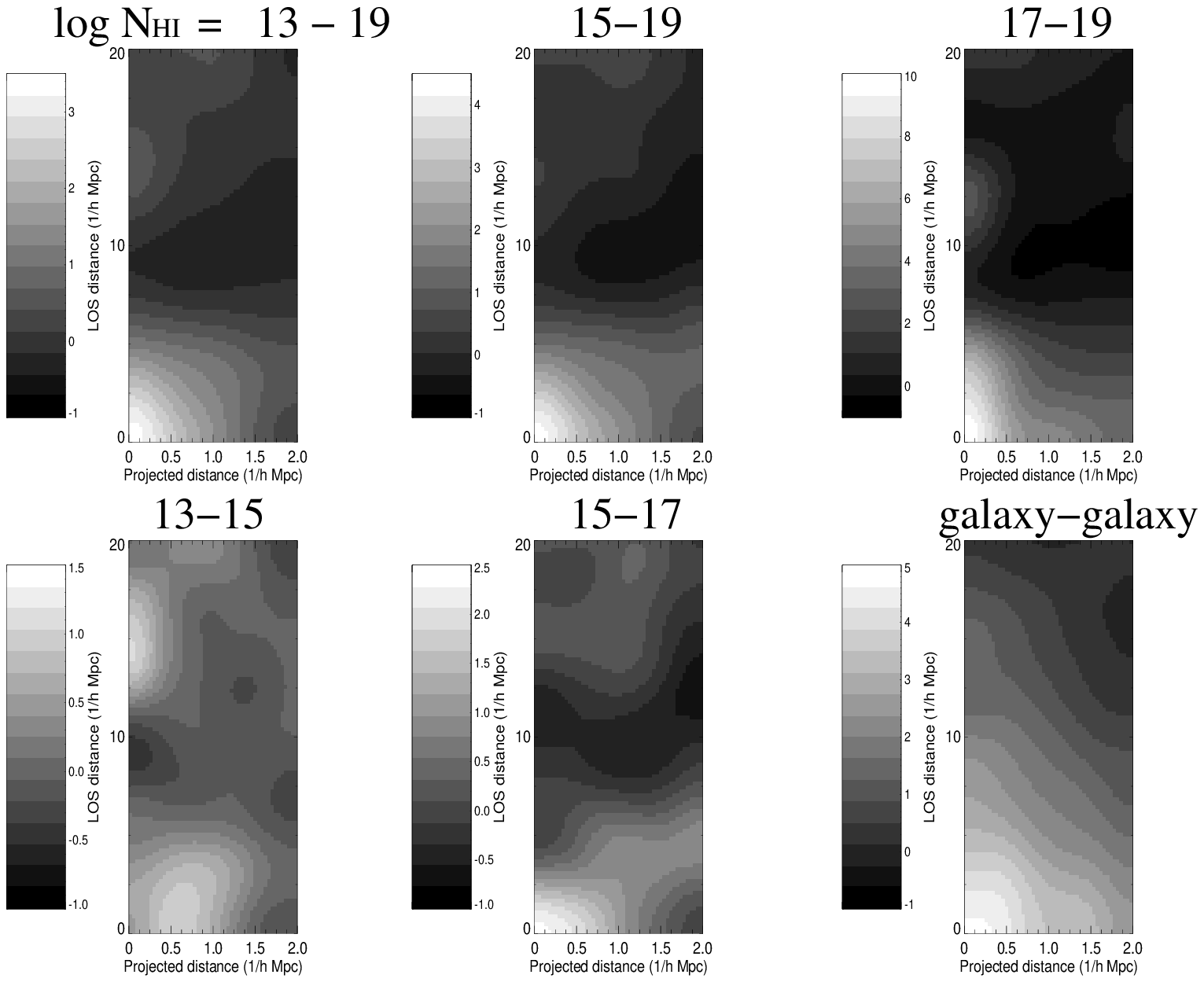}
\caption{\normalsize $\xi_{\rm{AG}}$ evaluated on a finer grid than shown in Fig.~\ref{fig:ALLbin} and smoothed as described in section 2.1. 
Results are presented for various intervals of HI absorber column
density (as indicated) and the galaxy-galaxy clustering is also
shown. Note the variations in greyscale value, and refer to the
estimated uncertainties in Table 2 when comparing the peak
$\xi_{\rm{AG}}$ values of the subpanels.}
\label{fig:ALLsm}
\end{figure*}

The evaluation of $\xi_{\rm{AG}}$ for various absorber sub-samples is
presented in Fig.~\ref{fig:ALLbin}, with the same binning scheme as
that used in Fig.~\ref{fig:xi13bin}. As the lower threshold on the HI
column density is increased from $10^{13}-10^{16}$\psqcm, the strength
of the central peak in $\xi_{\rm{AG}}$ also increases smoothly from
$\simeq 4.5$ to $\simeq 6.7$. At $N_{\rm{HI}} > 10^{17}$\psqcm~the
peak strength jumps dramatically to $\simeq 15$ although the formal
jackknife error is substantial. For absorber samples with a lower
maximum HI column density, the peak in $\xi_{\rm{AG}}$ is of lower
amplitude, $\simeq 3$. For the absorbers with $N_{\rm{HI}} =
10^{13}-10^{15}$\psqcm, the peak is spatially offset from the central
position. These results are summarised in Table~2 where we list the
peak values of $\xi_{\rm{AG}}$, their jackknife uncertainties, and
statistical significance assuming Poisson statistics. In
Fig.~\ref{fig:ALLsm} we show cross-correlation functions for the same
HI column density intervals evaluated on a fine grid but with the
underlying pair count fields heavily smoothed, as described in section
2.1. 

In Fig.~\ref{fig:ALLsm} we also show the auto-correlation function of
the galaxies in the sample. We note that whilst the peak value is
comparable to other measurements of the galaxy-galaxy 2-point
correlation function on scales of 1--2$h^{-1}$\Mpc~(e.g. from the 2DF;
Hawkins et al.~2003), it is formally lower than the peak in
$\xi_{\rm{AG}}$ for the $N_{\rm{HI}} = 10^{17}-10^{19}$\psqcm~sample, which
appears surprising. However, in view of the substantial uncertainty on the
peak value of $\xi_{\rm{AG}}$ for the latter subsample (see Table 2),
any such differences in peak amplitude are not significant. We also note
that other authors have compared the absorber-galaxy cross-correlation 
function to the galaxy-galaxy auto-correlation function to estimate
the mass of the dark matter haloes in which the absorbers are
embedded. Whilst this procedure is acceptable for Damped Lyman Alpha
absorbers (which are likely to be galaxies) (e.g. Bouch\'{e} et
al.~2004), it is not appropriate for diffuse Ly$\alpha$ absorbers
below $N_{\rm{HI}} = 10^{15}$\psqcm. The latter are likely to arise in 
unbound cosmological filaments (e.g. Dav\'{e} et al. 1999) so we do
not pursue such calculations here. Indeed, Ryan-Weber (2006) used this 
method and found that weak Ly$\alpha$ absorbers should
arise in $10^{14}$\Msun~groups; this seems unphysical since gas in
such groups would be quite hot, while the individual galaxies'
interstellar media do not have nearly the cross-section to reproduce
the counting statistics of weaker \Lya~forest absorbers.  

For CIV absorbers, Fig.~\ref{fig:ALLbin} shows that the peak strength of $\xi_{\rm{AG}}$ is comparable to that of HI absorbers with $N_{\rm{HI}} > 10^{16.5}$\psqcm. This is not surprising as in our dataset the median HI column density of CIV absorbers is $10^{17.7}$\psqcm, while the median of the full Ly$\alpha$ absorber sample is $10^{14.2}$\psqcm. The median HI column density of Ly$\alpha$ absorbers with $N_{\rm{HI}} > 10^{16.5}$\psqcm~is $10^{17.45}$\psqcm.


\begin{table}
\caption{Distribution of HI absorber column density}
\begin{tabular}{|ll|} \hline
log $N_{\rm{HI}}$ $\dagger  $  & Number of absorbers \\ \hline
$>13$ & 379 \\
$>14$ & 241 \\
$>15$ & 161 \\
$>16$ & 91 \\
$>17$ & 39 \\
$>18$ & 16 \\ \hline
\end{tabular} \\
$\dagger$ Inferred assuming Doppler parameter $b =30$\kmps. 
\end{table}

\begin{table}
\caption{Peak strength of $\xi_{\rm{AG}}$ shown in plots of Fig.~\ref{fig:ALLbin}}
\begin{tabular}{|llll|} \hline
log $N_{\rm{HI}}$  & Peak $\xi_{\rm{AG}}$ & Peak pair            & Peak \\ 
                   &                      & real (random) counts$^{\star}$   & significance $^{\star \star}$    \\ \hline
13-19 &  $4.4 \pm 1.7$ & 13 (2.4) & $2 \times 10^{-6}$\\
14-19 & $5.2 \pm 2.5$ &  13 (2.1) & $4 \times 10^{-7}$ \\
15-19 &  $6.0 \pm 3.8$ & 13 (1.9) & $1 \times 10^{-7}$ \\
16-19 & $6.7 \pm 5.2$ & 9 (1.2) & $4 \times 10^{-6}$ \\
17-19 & $15 \pm 12$ &  7 (0.44) & $4 \times 10^{-7}$ \\
13-15 $\dagger$ &  $3.0 \pm 2.0$ & 5 (1.3) & $9 \times 10^{-3}$ \\
15-17 & $3.3 \pm 1.1$ & 6 (1.4) & $3 \times 10^{-3}$ \\ 
CIV systems & $11 \pm 9.9 $ & 11 (0.94) & $1 \times 10^{-8}$ \\ \hline
\end{tabular} \\
$\star$ Number of real (random) absorber-galaxy pairs in cell at $\xi_{\rm{AG}}$ peak \\
$\star \star$ Poisson probability of observing $\geq $ peak pair
counts. These are included for completeness but substantially
overestimate the significance level of any signal; the
jackknife errors offer a more reliable significance measure. \\
$\dagger$ Peak not in central bin
\end{table}


\subsection{Dependence of $\xi_{\rm{AG}}$ on galaxy spectral type}
Chen et al.~(2005) reported that the strength of the absorber-galaxy cross-correlation may be dependent on galaxy spectral type, in the sense that a strong correlation signal is found for emission-line galaxies but not for absorption-line galaxies. Here we investigate this issue by splitting our galaxy sample into two sub-samples, according to whether the redshift assigned in paper I was obtained by identification of a strong emission line or where no such emission line could be seen. The resulting emission-line and absorption-line galaxy samples contain 406 and 225 galaxies, respectively. Of the original full galaxy sample of 685, a total of 54 galaxies could not be classified and were not included in this analysis: 8 of them have spectra with an ambiguous classification, and the redshifts of 46 galaxies were obtained from the literature without associated spectral information (see paper I for full details). The 225 absorption-line galaxies include most of the 49 very low redshift objects which were classified as probable stars in paper I, but this is not critical because such objects are too low in redshift to overlap with the absorption line sample.

\begin{figure*}
\includegraphics[width=0.7\textwidth,angle=0]{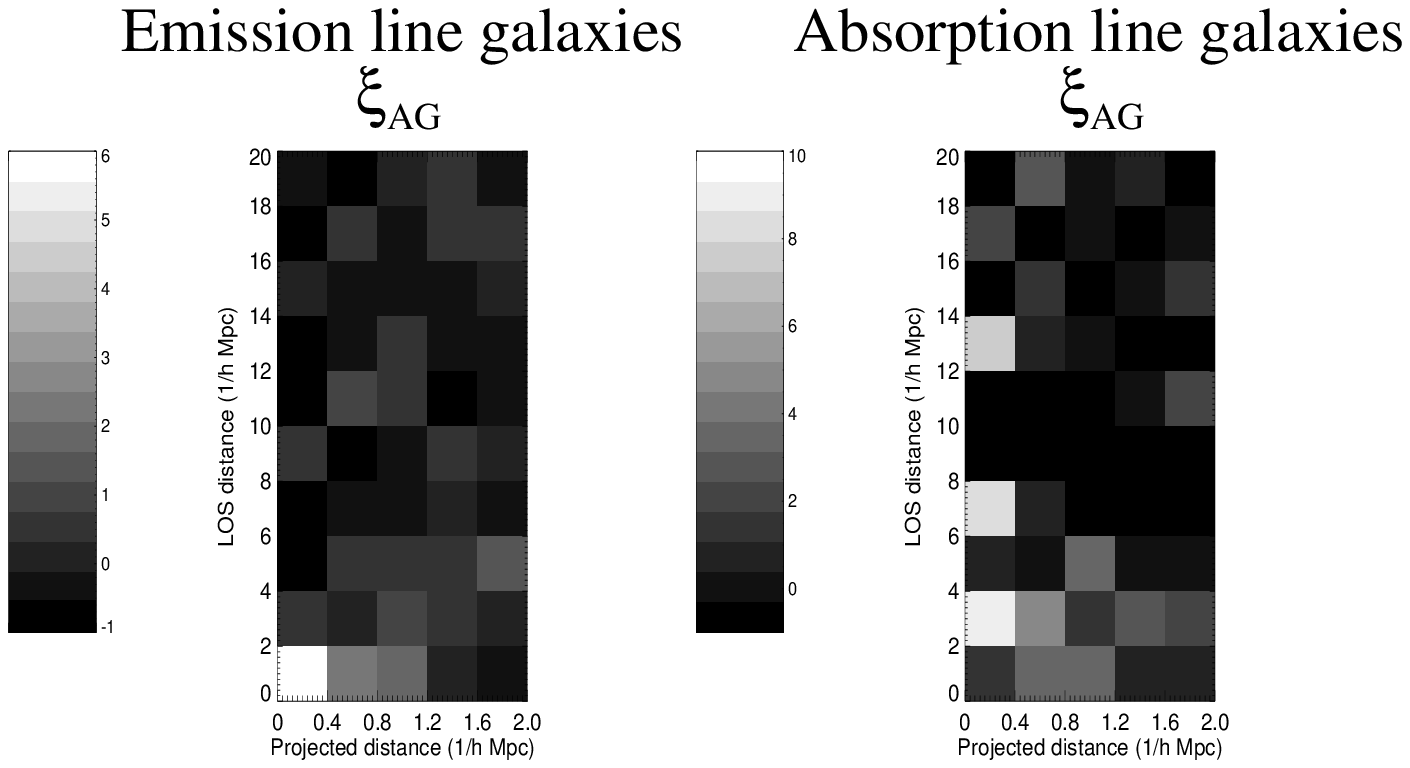}
\caption{\normalsize $\xi_{\rm{AG}}$ evaluated for the emission and absorption-line galaxy sub-samples using the full sample of absorption lines with $N_{\rm{HI}} > 10^{13}$\psqcm, as described in section 3.2}
\label{fig:galsplit}
\end{figure*}

These two galaxy samples were cross-correlated with the full HI absorption line catalogue (i.e. $N_{\rm{HI}} > 10^{13}$\psqcm) and the results are displayed in Fig.~\ref{fig:galsplit}. For the emission-line sample, the form of $\xi_{\rm{AG}}$ is qualitatively similar to the result obtained with the full galaxy sample (Fig.~\ref{fig:xi13bin}), but the peak strength is now marginally (but not significantly) higher at $\simeq 5.8$, compared with $4.4 \pm 1.7$ for the full galaxy sample. With just the absorption-line galaxies, the central peak in $\xi_{\rm{AG}}$ is observed to be displaced from the central bin to 2--4~$h^{-1}$\Mpc~along the LOS, although we do not claim that this is necessarily a statistically significant finding, in view of the much smaller size of the absorption-line galaxy sample. A more direct comparison with the 1-D results of Chen et al.~is presented in the next subsection. 



\begin{figure}
\includegraphics[width=8.5cm,angle=0]{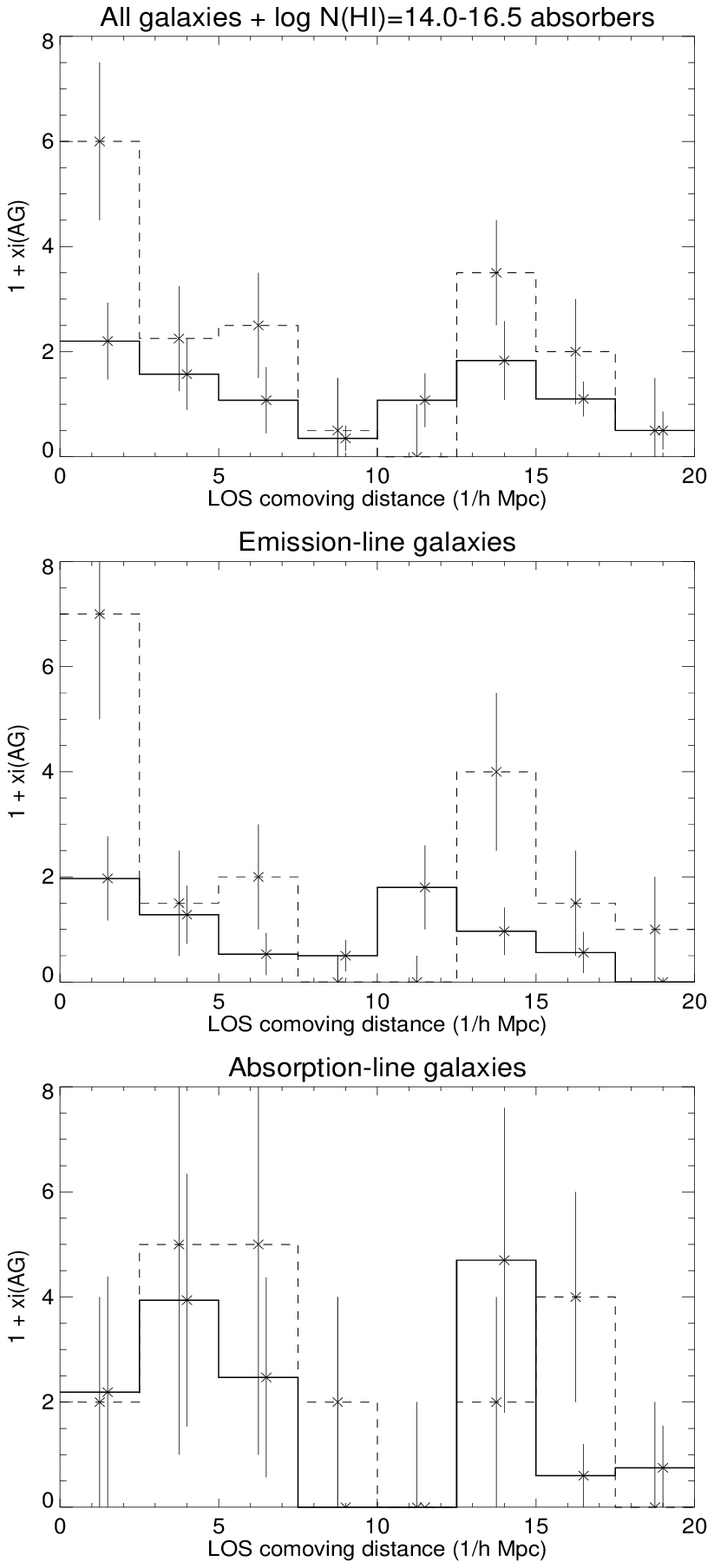}
\caption{\normalsize A comparison of our $\xi_{\rm{AG}}$ measurements (thick lines) with those of Chen et al.~(2005) (dashed lines) for different galaxy subsamples and absorption lines with $10^{14} < N_{\rm{HI}} < 10^{16.5}$\psqcm (upper panel: all galaxies: middle panel: emission-line galaxies; lower panel: absorption-line galaxies). $1 + \xi_{\rm{AG}}$ is plotted as a function of comoving line-of-sight distance for galaxies within $1 h^{-1}$\Mpc~of the line of sight. Error bars on our data points have been displaced in the x-direction for clarity.}
\label{fig:Chencomp}
\end{figure}

\subsection{Comparison with other works}
Of the $\xi_{\rm{AG}}$ measurements mentioned in the Introduction, that by Chen et al.~(2005) is the most similar to our work in its methodology, as it is based on an optical magnitude-limited galaxy redshift survey along the LOS to a single quasar (PKS 0405-123 at $z=0.5726$). Here we make a comparison between our results and their 1-D measurements of $\xi_{\rm{AG}}$ as function of galaxy type. We begin with a summary of the Chen et al. dataset. Their galaxy sample comprises 482 galaxies with $R \leq 20$~mag and spectroscopic redshifts over a 1600~arcmin$^{2}$ area within a $5h^{-1}$\Mpc~(comoving) impact parameter of the quasar LOS; the redshift distribution has a broad peak at $z \simeq 0.2$. For their $\xi_{\rm{AG}}$ analysis they restrict their attention to galaxies with impact parameter $<1 h^{-1}$\Mpc~of the LOS, of which there are 61 (46 emission-line galaxies, and 15 absorption-line galaxies; no details of the spectral classification method are given). The redshift distribution of this small impact parameter subsample is skewed towards lower redshifts compared with the entire sample, with a median redshift $z \simeq 0.1$. They claim a galaxy redshift accuracy of $\Delta z = \pm 0.0002$, or $\sim 40$\kmps. 

The Chen et al. absorption line sample comprises 112 Ly$\alpha$ systems at redshifts $0.01 \leq z \leq 0.557$ with HI column densities $log N_{\rm{HI}} = 12.5-16.5$ (80 per cent of the lines are at $log N_{\rm{HI}} < 13.6$, the regime in which our absorption line sample is most incomplete). The bulk of the lines were derived from observations with the {\em HST} Space Telescope Imaging Spectrograph (STIS) with spectral resolution of $6.7$\kmps. The latter is substantially higher than the $\sim 230$\kmps~resolution of the FOS spectra from which our absorption line sample was derived. Compared with our dataset, the Chen et al. sample can thus in principle probe $\xi_{\rm{AG}}$ on (i) smaller velocity scales; (ii) for weaker absorption lines ; (iii) with fewer complications arising from absorption line blending. The first two of these differences can be handled by comparing on appropriate $\Delta v$ scales and for sufficiently high $N_{\rm{HI}}$, but the issue of line blending in the FOS spectra is less straightforward and will be discussed later.

Mindful of the above differences between the datasets, we restrict our 
attention to a comparison with the Chen et al. $\xi_{\rm{AG}}$ measurements 
for absorption lines with $log N_{\rm{HI}} =14.0-16.5$. 
Chen et al. computed $\xi_{\rm{AG}}$ for galaxies within $1 h^{-1}$\Mpc~comoving of the LOS for 
the entire galaxy sample and its emission-line and absorption-line galaxy 
subsamples, computed out to LOS velocity separations of $\Delta v = 5000$\kmps. 
Converting $\Delta v$ to comoving distance, we plot the Chen et al. 
measurements along with our results out to LOS separations of 20$h^{-1}$\Mpc~in 
Fig.~\ref{fig:Chencomp}. Error bars on our data points were derived using 
jackknife resampling; Chen et al. give no details of their error bar computation.  

The plots in Fig.~\ref{fig:Chencomp} demonstrate that we do not
reproduce the high peak in $\xi_{\rm{AG}}$ at the smallest separations
(comoving LOS separations $<2 h^{-1}$\Mpc) found by Chen et al. for
the entire galaxy sample: our peak values are more than $2 \sigma$
apart. Possibly related to this, we also do not substantiate their 
finding of much stronger $\xi_{\rm{AG}}$ for the emission-line
galaxies as compared with the absorption-line galaxies. Whilst our
dataset is substantially larger than that of Chen et al. in terms of
the numbers of galaxies (our
study: 179 galaxies with impact parameters $<1 h^{-1}$\Mpc~overlapping
with the absorption-line redshift distribution; Chen et al: 61
galaxies) and absorbers (our study: 180; Chen et al.: 15), a much
larger dataset will be needed to address more conclusively any
dependence of $\xi_{\rm{AG}}$ on galaxy type. It should also be borne
in mind that the Chen et al. study is effectively at redshift $z \sim
0.1$ (in terms of maximal overlap between absorber and galaxy
samples), whereas our study is at $z \sim 0.5$. Concerning the
possible impact of different degrees of absorption-line blending in
quasar spectra from the FOS and STIS, we refer to Penton, Stocke \&
Shull~(2004): they smoothed their STIS/Goddard High Resolution
Spectrograph (GHRS) spectra to FOS resolution and found a 15--25~per
cent increase in the line density at REW$>0.24$\AA~($N_{\rm{HI}}
\simeq 10^{14}$\psqcm~for $b=30$\kmps) due to the blending of weaker
lines. Since Chen et al. found that these weaker absorption lines
correlate much less strongly with galaxies, this alone would suppress
our peak $\xi_{\rm{AG}}$ values by 25--35~per cent relative to their
measurements, going some way to explaining the differences in
Fig.~\ref{fig:Chencomp}. However, the full impact on the
$\xi_{\rm{AG}}$ measurements of differences between our respective galaxy and
absorption-line samples merits further investigation. Finally we note 
that the 1-D $\xi_{\rm{AG}}$ in Fig.~\ref{fig:Chencomp} all exhibit 
possible minima at LOS separations $\simeq 10 h^{-1}$\Mpc, similar to 
the feature in the smoothed 2-D plots of Fig.~\ref{fig:ALLsm}; the 
physical origin of this feature is not known.

Ryan-Weber~(2006) cross-correlated quasar absorption line systems with
$12.4 \leq log N_{\rm{HI}} \leq 14.8$ against an HI-selected galaxy
catalogue from the HIPASS survey. As in the Chen et al. absorption
line sample, this absorption line sample is based on {\em HST} STIS
and GHRS spectra and thus provides considerably higher velocity
resolution and weak line sensitivity compared with our sample. The
galaxy sample is biased towards gas-rich galaxies, including gas-rich 
dwarf and low surface brightness galaxies often absent from
magnitude-limited optical catalogues such as ours. These differences, together with
the lower redshift of her measurements (which are effectively at $z
\simeq 0$), might account for the extreme LOS elongation seen in her $\xi_{\rm{AG}}$ measurements -- out to $10 h^{-1}$\Mpc~comoving -- which we do not reproduce. Indeed, morphological analysis of the optical counterparts to the 1000 brightest HIPASS sources confirms that the majority are late-type gas-rich galaxies (Zwaan et al.~2003). The origins of this LOS elongation are explored further in section 4.2 below with the help of SPH simulations. In seeking to compare our $\xi_{\rm{AG}}$ results with those of Ryan-Weber, an important first step is to compare the ranges of galaxy mass probed by the two surveys. Ryan-Weber states that the geometric mean HI mass of galaxies contributing to absorber-galaxy pairs is $log (M_{\rm{HI}}/\Msun) = 8.8 h^{-2}$, with a corresponding dark halo mass of $log (M/\Msun) = 11.0 h^{-1}$ (as inferred from modelling by Mo et al.~2005). In contrast, our galaxy sample was selected by R-band imaging and follow-up spectroscopy and is thus expected to contain a higher fraction of early-type galaxies. In paper I (section 2.3.5) we attempted to place bounds on the rest-frame B-band absolute magnitudes of our galaxy sample: confining our attention to galaxies at redshifts $0.35 \leq z \leq 0.8$ which overlap with the absorption line sample, we find that the vast majority of these galaxies have $-22 \leq M_{\rm{B}} \leq -18$~mag, with a median $M_{\rm{B}} \simeq -20.5$~mag. With reference to models of the conditional probability distribution of halo masses (e.g. Cooray 2005 and references therein), we expect galaxies at this median luminosity to be hosted by dark matter haloes of mass $\sim 10^{12-13} h^{-1}$\Msun.

We note that not only are our galaxies hosted by more massive haloes than those of Ryan-Weber, but that the two samples fall either side of a critical halo mass, at which the nature of gas accretion onto galaxies is thought to change. Using SPH simulations, Keres et al.~(2005) find that galaxies hosted by haloes with $M_{\rm{halo}} \leq 10^{11.4}$\Msun~(or, equivalently, baryonic mass $M_{\rm{gal}} \leq 10^{10.3}$\Msun) acquire most of their gas via 'cold mode' accretion in which gas at $T < 10^{5}$K is efficiently drawn in from large distances along filaments. For more massive galaxies, the dominant mechanism is quasi-spherical accretion of hot shocked gas at the virial temperature of the halo, more akin to the classical picture outlined by White \& Rees~(1978). Moreover, this transition in accretion mode is a strong function of halo mass over the range of interest: from $M_{\rm{halo}}=10^{11}$\Msun~to $10^{12}$\Msun, the fraction of gas accreted by the `cold mode' accretion varies from $>0.9$ to $<0.1$, virtually independent of redshift. Similar results have been found by Dekel \& Birnboim~(2006). We do not wish to claim that this switch from cold mode filamentary accretion in Ryan-Weber's HIPASS galaxies to hot mode accretion in our galaxies can wholly explain the different degrees of LOS elongation in $\xi_{\rm{AG}}$ measurements using the two samples, but it may be a useful signpost to fruitful areas of future investigation.

\section{COMPARISON WITH SPH SIMULATIONS}
In this section we use cosmological simulations to generate synthetic 2D absorber-galaxy correlation functions to compare with the observations, and to gain insight into the physical origin of features such as the `finger of god' which is seen in the results of Ryan-Weber (2006). 

We ran a simulation using Gadget-2 (Springel \& Hernquist 2003)
with improvements as described in Oppenheimer \& Dav\'e (2006), in
a cubic comoving volume of 64$h^{-1}$Mpc on a side.  We employed the
momentum-driven outflow model `vzw' as described in Oppenheimer \&
Dav\'e , and a cosmology concordant with 3rd-year WMAP results (Spergel
et al. 2006): $\Omega_m=0.26$, $\Omega_\Lambda=0.74$, $H_0=71$~km/s/Mpc,
$\sigma_8=0.75$, and $\Omega_b=0.044$.  The force resolution is
$5h^{-1}$kpc, and the mass per gas particle is $2.72\times 10^8M_\odot$.
The galaxy finding procedure is described in Finlator et al. (2006), using a
minimum resolved galaxy stellar mass of $8.7\times 10^9 M_\odot$.

The simulation volume was pierced by 1000 randomly-located sightlines to simulate quasar spectra, and absorption line profiles 
were fitted using AutoVP (Dav\'e  et al.~1997). Absorber-galaxy pairs were measured and tabulated from these spectra for LOS separations $\Delta v < 1600$\kmps~and projected separations of $\Delta r < 16$\Mpc~(physical). Following conversion of pair separations into comoving redshift space, these catalogues were used to create 2-D absorber-galaxy correlation functions. Two such absorber-galaxy pair catalogues were created in order to compare with our observational results and those of Ryan-Weber, as we now describe. 

\subsection{Tuning the simulations to compare with the observational results}

Firstly, to compare with the observational results presented in section 3 of this paper, the simulation output was extracted at $z=0.5$ (giving a total redshift pathlength of $\Delta z =  26.4$). The pixel scale of the quasar spectra was set to 70\kmps~and convolved with a Gaussian of 230\kmps~FWHM, in order to match the resolution of the {\em HST} Faint Object Spectrograph. HI absorption lines were selected with rest equivalent width REW $>0.056$\AA, corresponding to column densities $N_{\rm{HI}} > 10^{13}$\psqcm~for $b=30 $\kmps. The strength of the ionizing background radiation was increased by a factor of 4 over the Haardt \& Madau~(1996) value in order to normalise the $dN/dz$ value for $REW > 0.24$\AA~lines to the HST Absorption Line Key Project results (Weymann et al.~1998). The entire resolved galaxy population was used for the comparison, with a galaxy density of 0.01145 per cubic $h^{-1}$\Mpc~comoving above the $8.7\times 10^9 M_\odot$ baryonic mass limit. As shown by, for example, the Deep Evolutionary Exploratory Probe 2 (DEEP2) Galaxy Redshift Survey (Willmer et al.~2006), this is comparable to the space density of galaxies at this redshift in an absolute range which matches our observed galaxy sample ($-22 \leq M_{\rm{B}} \leq -18$~mag; see section 3).

The second absorber-galaxy pair catalogue was extracted from the simulations at redshift $z=0$ (giving a total redshift pathlength of $\Delta z = 21.3$) to compare with the Ryan-Weber results. The spectral resolution of the artifical quasar sightlines was set to 8\kmps~to mimic the higher resolution {\em HST} STIS and GHRS spectra, and HI absorption lines were extracted for REW=0.0126--0.45\AA, corresponding to $10^{12.4} < N_{\rm{HI}} < 10^{14.8}$\psqcm~for $b=30$\kmps. Mimicing the HIPASS selection of galaxies in HI emission by means of a simple cut in baryonic mass is a crude approximation, but it is the only tool at our disposal with the current simulation. Somewhat arbitrarily, we use an upper baryonic mass cut-off for galaxies of $2 \times 10^{10}$\Msun. To put this figure in context, we refer to some results from Durham semi-analytic models of galaxy formation (Cedric Lacey, private communication; for the latest implementation of such models see Bower et al.~2006). Such models track the behaviour of gas and stars in dark matter haloes with simple physical prescriptions. At redshift $z=0$, dark halo masses of $10^{11}, 10^{12}, 10^{13}$ and $10^{14} h^{-1}$\Msun~contain galaxies of total baryonic mass (hot gas, cold gas and stars) of $10^{10}, 8.6 \times 10^{10}, 10^{12}$ and $10^{13}$\Msun, respectively. The corresponding cold gas masses are $1.7 \times 10^{9}, 6.7 \times 10^{9}, 3.3 \times 10^{10}$ and $2.7 \times 10^{11}$\Msun, respectively. A baryonic mass cut off of $2 \times 10^{10}$\Msun~corresponds to a cold gas mass of $4.2 \times 10^{9}$\Msun~and a dark halo mass of $2.5 \times 10^{11} h^{-1}$\Msun. As stated in section 3, the HIPASS galaxies contributing to Ryan-Weber's absorber-galaxy pairs have mean HI and dark halo masses of $10^{8.8}/h^{2}$ and $10^{11}/h$\Msun, respectively. The space density of these simulated galaxies is 0.006107 per cubic $h^{-1}$\Mpc~comoving.

\begin{figure*}
\includegraphics[width=0.84\textwidth,angle=0]{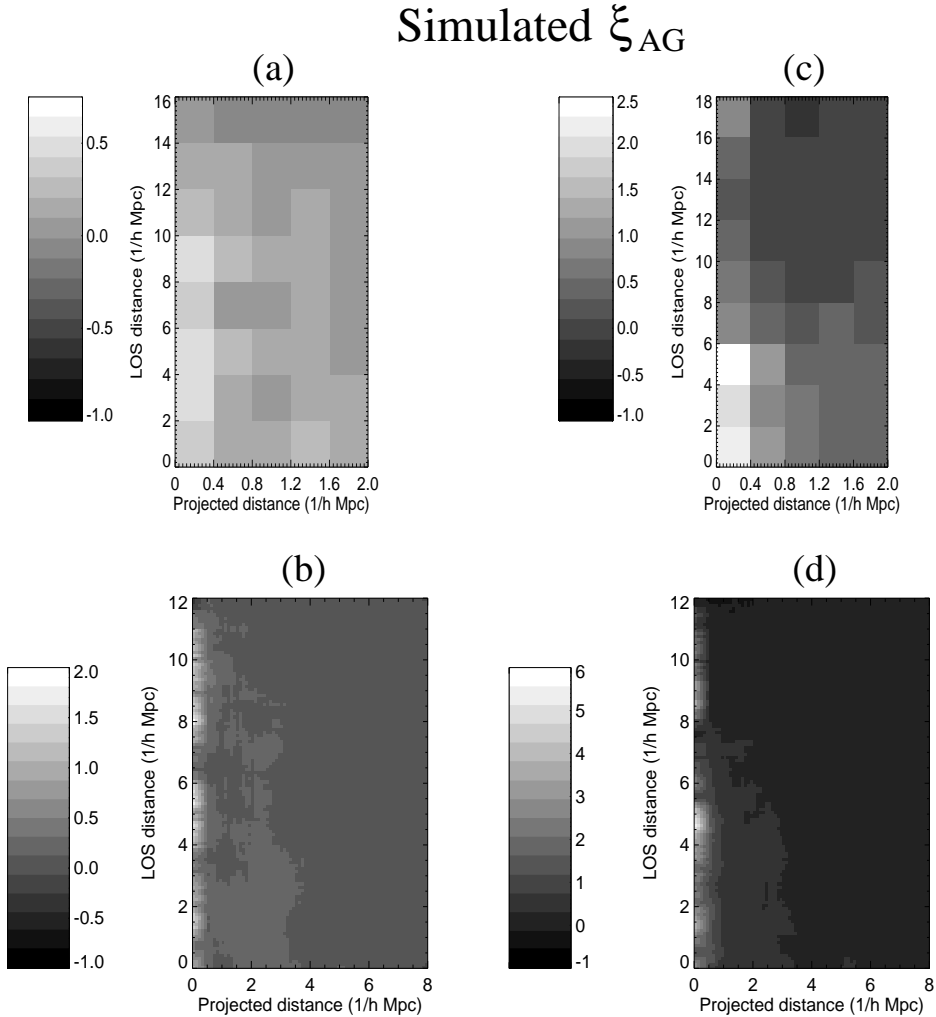}
\caption{\normalsize $\xi_{\rm{AG}}$ for the two simulations described in section 4.1, each plotted in two different ways. (a) The simulation with STIS-resolution sightlines, cross-correlating $10^{12.4} < N_{\rm{HI}} < 10^{14.8}$\psqcm~absorbers with galaxies of baryonic mass $<2 \times 10^{10}$\Msun~at $z=0$. (b) As in (a), but binned/smoothed as in Ryan-Weber~(2006). (c) The simulation with FOS-resolution sightlines, cross-correlating $N_{\rm{HI}} > 10^{13}$\psqcm~absorbers at $z=0.5$ with the entire resolved galaxy population; (d) as in (c) but binned/smoothed as in Ryan-Weber~(2006).}
\label{fig:simulated}
\end{figure*}


\subsection{Results: the simulated $\xi_{\rm{AG}}$ and the origin of the `finger of god'}
The two-dimensional $\xi_{\rm{AG}}$ which result from the simulations described in section 4.1 are displayed in Fig.~\ref{fig:simulated}. Plots with two types of binning/smoothing are presented, firstly binned coarsely to match our results in Fig.~\ref{fig:ALLbin} (although the simulated plots extend out to slightly smaller LOS distances), and secondly using the technique of Ryan-Weber. The latter evaluates $\xi_{\rm{AG}}$ on a fine grid ($0.1 \times 0.1 h^{-1}$\Mpc) and then smoothes the resulting $\xi_{\rm{AG}}$ by $9 \times 9$ pixels. Note that this differs from the smoothing procedure we used to generate Fig.~\ref{fig:ALLsm}, where the underlying pair counts were smoothed before computing $\xi_{\rm{AG}}$.

Comparing the simulations of Fig.~\ref{fig:simulated} with our observational results in Fig.~\ref{fig:ALLbin} and with the results of Ryan-Weber, we notice that both sets of observed $\xi_{\rm{AG}}$ are much more strongly peaked at the smallest separations than the simulations, i.e. we observe much more gas at small velocity offset and/or close to the galaxies than present in the simulations. Redshift measurement errors in the observations would tend to weaken such peaks, so this cannot be the origin of the discrepancy. A related issue is the asymmetric elongation of the simulated $\xi_{\rm{AG}}$ in redshift space along the LOS, which is qualitatively similar to that found by Ryan-Weber but which appears to be absent from our measurements. Such a feature is analogous to that seen in the autocorrelation function of galaxies (see e.g. Hawkins et al.~2003), where it arises from departures from Hubble flow due to virialised motions in groups and clusters. Commonly referred to as the `finger of god', the effect was seen in $\xi_{\rm{AG}}$ in an earlier generation of cosmological simulations (e.g. Dav\'{e} et al.~1999) but its physical interpretation is not as clear as in the galaxy-galaxy case. Ryan-Weber~(2006) interpreted it as arising from the draining of gas from low-density regions into collapsed structures. In a subsequent paper we will address the role of galaxy outflows in causing the `finger of god', and also investigate the dependence of $\xi_{\rm{AG}}$ on galaxy mass and absorption line column density. A more ambitious goal is to simulate the full observational dataset, including the selection of the galaxies in a somewhat more realistic manner than is possible with the existing cuts in total baryonic mass.

In closing this section, we comment on another possible explanation for the `finger of god' which merits further investigation. It may be a geometrical effect which leads to a bias towards observing large numbers of galaxy-absorber pairs at small projected sky separations when a LOS is oriented directly down an IGM filament with clumpy HI absorption. At redshifts $z \leq 2$, simulations show that absorbing gas with $N_{\rm{HI}} < 10^{15}$\psqcm~resides in a filamentary network (e.g. Dav\'{e} et al.~1999), whereas at higher redshifts such gas is distributed more diffusely. This basic structure evolves little over $0 \leq  z  \leq 2$ but absorbing material of a given column density becomes increasingly clumpy towards lower redshifts. This might partially account for the increased LOS elongation measured by Ryan-Weber at $z \simeq 0$ compared with our measurements at $z \simeq 0.5$, but further investigations are needed. If only a very small proportion of sightlines are fortuitously aligned with filaments, it is much more likely that such chance alignments occur in the simulations (1000 LOS; $\Delta z \sim 25$) which have a much higher redshift pathlength than our dataset (16 LOS; $\Delta z \sim 3$) or that of Ryan-Weber (27 LOS; $\Delta z  \sim 1$).

\section{CONCLUSIONS}
In summary, our measurements show that the strength of the correlation function between HI quasar absorption systems and galaxies increases with HI column density, certainly for comoving separations $<0.4 h^{-1}$\Mpc~(projected) and $<2 h^{-1}$\Mpc~(line-of-sight) where the signal:noise of our measurements is greatest.  The peak strength of the absorber-galaxy clustering increases only slowly as the lower $N_{\rm{HI}}$ limit is increased from $10^{13}$ to 
$10^{16}$\psqcm, but then jumps considerably (albeit with substantial uncertainty) above $10^{17}$\psqcm. If real, such a jump may reflect the transition from absorption 
within the filamentary IGM to absorption within the haloes of individual galaxies. The peak strength of the CIV absorber-galaxy cross-correlation is comparable to that of the HI absorber-galaxy signal for $N_{\rm{HI}} > 10^{16.5}$\psqcm.

Our results are broadly consistent with the finding of Chen et
al.~(2005), based on the sightline to the quasar PKS 0405-123, that
the absorber-galaxy cross-correlation strength is insensitive to HI
column density in the range $10^{13.6} \leq N_{\rm{HI}} <
10^{16.5}$\psqcm, although Table~2 shows evidence for a weak increase
with $N_{\rm{HI}}$. We find tentative but inconclusive evidence for
differences in the 2-D absorber-galaxy cross correlation function for
our subsamples of emission-line and absorption-line galaxies, but we
do not substantiate the claim of Chen et al. that the 1-D cross-correlation signal is much higher for emission-line galaxies than for absorption-line galaxies. This discrepancy between our results and those of Chen et al. is not fully understood, but it may be partially ascribed to the effects of line blending in the FOS spectra compared with the higher resolution STIS quasar spectra.

Comparison was also made with the results of Ryan-Weber~(2006) on the cross-correlation of HI-selected galaxies with quasar absorbers with $10^{12.4} \leq N_{\rm{HI}} < 10^{14.8}$\psqcm, who find a marked `finger of god' elongation out to $10 h^{-1}$\Mpc~along the sight. Such a feature appears to be absent from our measurements, but we demonstrated that there are significant differences in the characteristic mass of our respective galaxy samples which may account for such differences. The dark matter halo masses of the two samples fall either side of a critical dark halo mass of $10^{11.4}$\Msun~(with the Ryan-Weber sample being lower in mass), at which simulations show that the nature of the dominant mode of gas accretion onto galaxies changes fundamentally from cold filamentary accretion in low mass haloes to hot quasi-spherical accretion (Keres et al.~2005).

We also constructed synthetic absorber-galaxy 2-point correlation functions using output from a $64 h^{-1}$\Mpc~box SPH simulation aimed at reproducing our own observational results and those of Ryan-Weber~(2006). The simulations revealed somewhat less gas at the smallest separations from galaxies than present in the observations and a strong `finger of god' elongation of $\xi_{\rm{AG}}$ along the LOS. These discrepancies may originate in true deficiencies in the physics of the SPH simulations, or in a failure to accurately mimic the observed galaxy selection functions, or through some more subtle selection effect. 

For future work, we suggest that effort be invested to understand the dominant contribution to the `finger of god' elongation in $\xi_{\rm{AG}}$, and the nature of the apparent transition in $\xi_{\rm{AG}}$ between $N_{\rm{HI}}=10^{16}$ and $10^{17}$\psqcm. On the theoretical side, comparison should be made with a variety SPH simulations matching the observational selection function more closely. On the observational side, it would also be valuable to conduct surveys for higher-redshift galaxies around existing HST Quasar Absorption Line Key Project sightlines to ensure a better match between the redshift distributions of the absorption lines and the galaxies, and around higher-resolution HST STIS quasar sightlines. It would also be valuable to survey some of these sightlines with large-format IFUs to enhance sensitivity to emission-line galaxies which may be preferentially missed by our existing broad-band optical selection techniques.

\section*{ACKNOWLEDGMENTS}
This work has been partially financed by a Rolling Grant at Durham
University from the UK Particle Physics and Astronomy Research
Council. BTJ acknowledges partial support from NASA through
HF-1045.01-93A and from the National Optical Astronomy Observatory,
operated by AURA, Inc., on behalf of the US National Science
Foundation. Support for RD was provided in part by NASA through grant
HST-AR-10308.01-A from the Space Telescope Science Institute, which is
operated by the Association of Universities for Research in Astronomy,
Inc., under NASA contract NAS5-26555. We thank Bob Carswell, Cedric
Lacey and Neil Crighton for useful conversations, and the anonymous
referee for an extensive and constructive report. RJW gratefully
acknowledges the hospitality of the Harvard-Smithsonian Center for
Astrophysics during the completion of this paper.

{}

\end{document}